\documentstyle[12pt]{article}
\def\Journal#1#2#3#4{{#1} {\bf #2}, #3 (#4)}

\def\be{\begin{equation}}
\def\ee{\end{equation}}
\def\bea{\begin{eqnarray}}
\def\eea{\end{eqnarray}}

\begin{document}

\begin{titlepage}
\thispagestyle{empty}
\begin{center}
{ \bf
STOCHASTIC REACTION-DIFFUSION PROCESSES, OPERATOR ALGEBRAS AND 
INTEGRABLE QUANTUM SPIN CHAINS\footnote{Invited lecture given at the
Satellite Meeting to Statphys 19 on Statistical Models, Yang-Baxter Equation
and Related Topics, at Nankai University, Tianjin (August 1995). 
To appear in the
Proceedings, eds. F.Y. Wu and M.L. Ge, (World Scientific, Singapore)}
}\\[25mm]

{ {\sc
Gunter M. Sch\"utz}
}\\[8mm]

\begin{minipage}[t]{10cm}
\begin{center}
{\small\sl
Department of Physics, University of Oxford\\
Theoretical Physics, 1 Keble Road, Oxford OX1 3NP, UK
}
\end{center}
\end{minipage}
\vspace{30mm}
\end{center}
{\small
We show that the stochastic dynamics of a large class of 
one-dimensional interacting particle systems may be presented by integrable 
quantum spin Hamiltonians. 
Using the Bethe ansatz and similarity transformations 
this yields new exact results. In a complementary approach we generalize 
previous work\cite{Sti1,Sti2} and present a new description of these and other
processes and the related quantum chains in terms of an operator algebra
with quadratic relations.
The full solution of the master equation of the process is thus
turned into the problem of finding representations of this algebra. We find
a two-dimensional time-dependent representation of the algebra for the 
symmetric exclusion process with open boundary conditions. We obtain new 
results on the dynamics of this system and on the eigenvectors and eigenvalues 
of the corresponding quantum spin chain, which is the isotropic 
Heisenberg ferromagnet with non-diagonal boundary fields.
}
\\
\vspace{5mm}\\
\end{titlepage}

\section{Stochastic Dynamics and Quantum Systems}

A convenient and much used description of stochastic processes is in terms
of a master equation for the probability distribution of the stochastic
variables of the system. A master equation expresses the probability of 
finding the system at time $t+\Delta t$ in a given configuration in terms of 
the probability distribution at time $t$ through a first order differential
equation in the time variable (infinitesimal $\Delta t$) or through a 
difference equation in $t$ (for discrete time steps $\Delta t$). Such 
processes are Markov processes which may be
constructed for the description of e.g. interacting particle systems
\cite{Ligget}. These systems are of interest because they have turned out
to be useful as models for reaction-diffusion systems in physics and chemistry
\cite{Priv} and, through various mappings, as models for spin flip dynamics 
\cite{Glauber}, interface growth \cite{KS}, dynamics of DNA in gels \cite{ZS} 
and many other highly interesting and non-trivial systems. Even in relatively 
simple models one finds a very rich dynamical 
behaviour involving dynamical and non-equilibrium phase transitions of
various kinds. A particularly well-studied model where this happens is the
asymmetric exclusion process in one dimension \cite{ZS}. This model is a 
discrete version of the noisy Burgers equation and describes not only driven 
diffusion of hard-core particles, but is also
a model for dynamics of driven interfaces, polymers in random media \cite{KS}
and the kinetics of biopolymerization \cite{bio}.

It is well-known that a master equation can be expressed in a ``quantum
Hamiltonian formalism'' by mapping each state of the system to a basis 
vector in a suitable vector space $X$. In this mapping the probability 
distribution at time $t$ becomes a vector $|\,f(t)\,\rangle$ and the master 
equation takes the form
\be
\frac{\partial}{\partial t} |\,f(t)\,\rangle = - H |\,f(t)\,\rangle
\label{1}
\ee
where $H$ is a suitably chosen linear map acting on $X$.\footnote{
For discrete time dynamics the corresponding equation reads
$|\,f(t+1)\,\rangle = T |\,f(t)\,\rangle$.} 
This is in formal analogy to second quantization in quantum mechanics and we 
will therefore, in slight abuse of language, call $H$ a quantum Hamiltonian. 
The ground state of this in general non-hermitian Hamiltonian
(which by construction has energy 0) corresponds to the stationary 
probability distribution of the stochastic dynamics.

This mapping is in itself not a great achievement, since it represents only
a change of language. However, it has turned out in recent
years that for many interesting models the quantum Hamiltonian obtained in 
this way is an object well-known from other areas of physics and tractable
with techniques largely unknown to the community of people interested in
the original problem of the stochastic dynamics. A paradigmatic example
of this kind is the symmetric exclusion process \cite{Spi}. In this model
particles on a lattice hop between lattice sites $k,l$ with rates $p(k,l)
=p(l,k)$. They interact via a hard-core repulsion which prevents the
occupation of a lattice site by more than one particle. 
The quantum Hamiltonian obtained for this system through the mapping described
above is the Hamiltonian for the isotropic spin 1/2 Heisenberg 
ferromagnet\cite{AH}. Here the $SU(2)$ symmetry of the problem (which is not 
obvious at all in the original master equation) and other 
approaches have been used for obtaining new exact results \cite{GS,scsa}. 
Moreover, in one dimension, the system with nearest neighbour hopping is 
integrable and can be solved by the Bethe ansatz. 

As will be shown below, the integrability is not a special feature
of the symmetric diffusion process alone. Also the more interesting and
much more challenging case of asymmetric diffusion in one dimension is
described by an integrable quantum Hamiltonian and, using the Bethe ansatz 
and related methods many new exact results have been obtained 
\cite{GS,asym,HS,Kim}. Moreover, it turns out that a 10-parameter class
of reaction-diffusion systems of identical particles\cite{sch2} and 
reaction-diffusion systems of non-identical particles \cite{Alcaraz,Dahmen} 
are described by integrable quantum chains.\footnote{For discrete-time 
dynamics with parallel sublattice updating
one gets as time evolution operator $T$ the transfer matrices of integrable
vertex models \protect\cite{KDN,sch}.}

Independently from these developments a new approach for the treatment of 
one-dimensional systems was introduced by a matrix description of the ground 
states of spin Hamiltonians \cite{Zit} and then of the stationary distribution 
of the asymmetric exclusion process with open boundary conditions 
\cite{DEHP,Sven,vladimir}. 
In this approach the ground state vector of the quantum Hamiltonian (i.e. the 
stationary distribution of the stochastic process) is expressed in terms of a 
matrix product measure and given by certain matrix elements of matrices which 
are representations of an algebra determined from the dynamics (i.e. the 
Hamiltonian) of the system (see below). Going further, a time-dependent matrix 
ansatz was introduced for the description of the {\em complete} dynamics of the 
(a)symmetric exclusion process with open boundary conditions\cite{Sti1} and 
used for the derivation of the spectrum in the symmetric case\cite{Sti2}. 
While for the calculation of correlation functions one needs a representation
of the dynamical algebra, for the derivation of the spectrum alone the algebra
itself is sufficient. In general, the spectrum contains already valuable
information about the decay of correlations to their stationary values:
If the Hamiltonian has an energy gap, the decay will, at late times, be
exponential, while for a continous gapless spectrum one expects algebraic decay.
Moreover, from a finite-size scaling analysis of the spectrum one may
find the dynamical exponent of the system.

In this talk I will first show (Sec. 2) how reaction-diffusion systems of 
identical hard-core particles are related to a generalized Heisenberg chain. 
Its spectrum can be obtained from the Bethe ansatz. This will be a simplified 
rederivation of some results obtained earlier\cite{sch2}. Then (Sec.~3) I will 
generalize the operator approach to the general reaction-diffusion 
problem of identical hard-core particles with nearest neighbour interaction in 
one dimension. Finally I will return to the symmetric diffusion process and
present a two-dimensional representation of the time-dependent operator algebra.
I would like to emphasize that most of the results presented in Sec.~3 are 
original.

\section{Integrable Reaction-Diffusion Processes}

We will consider stochastic reaction-diffusion processes of identical particles 
with hard-core repulsion moving on a ring with $L$ sites. Even though part of 
our approach generalizes to arbitrary lattices \cite{sch2} we will study here 
only one-dimensional systems with nearest neighbour interaction. The stochastic 
variables of the system are the occupation numbers $\underline{n} = \{ n_k \}$ 
where $n_k=0,1$ indicates whether site $1 \leq k \leq L$ in
the lattice is occupied or empty. At a given time $t$ the state of the system
is completely described by the probability distribution $f(\underline{n};t)$.
In this class of models there are ten possible reactions in addition to 
right and left hopping (diffusion), so altogether one has to specify 12
independent rates $a_{ij}\geq 0$ (Tab.~1). 

\begin{table}[t]
\caption{Bulk reaction and diffusion rates for nearest neighbour exclusion
processes of identical particles. The numbers $a_{ij}$ are the rate of change
of the occupation numbers $\{n_k,n_{k+1}\}$.}
\vspace{0.4cm}
\begin{center}
\begin{tabular}{|c|c|l||c|c|l|}
\hline
& & & & & \\
Process & Rate & & Process & Rate & \\ \hline
& & & & & \\
01 $\rightarrow$ 10 & $a_{32}$ & diffusion &
10 $\rightarrow$ 01 & $a_{23}$ & diffusion \\
\hline
& & & & & \\
11 $\rightarrow$ 01 & $a_{24}$ & coagulation &
01 $\rightarrow$ 11 & $a_{42}$ & decoagulation \\
11 $\rightarrow$ 10 & $a_{34}$ & coagulation &
10 $\rightarrow$ 11 & $a_{43}$ & decoagulation \\
\hline
& & & & & \\
00 $\rightarrow$ 01 & $a_{21}$ & creation &
01 $\rightarrow$ 00 & $a_{12}$ & annihilation \\
00 $\rightarrow$ 10 & $a_{31}$ & creation &
10 $\rightarrow$ 00 & $a_{13}$ & annihilation \\
00 $\rightarrow$ 11 & $a_{41}$ & pair creation &
11 $\rightarrow$ 00 & $a_{14}$ & pair annihilation \\
& & & & & \\ \hline
\end{tabular}
\end{center}
\end{table}

The stochastic dynamics are defined by the master equation
\be
\frac{d}{dt} f(\underline{n};t) = \sum_{\underline{n}'}
\left[ w(\underline{n};\underline{n}') f(\underline{n}';t) -
w(\underline{n}';\underline{n}) f(\underline{n};t) \right] 
\ee
where the reaction-diffusion rates $w(\underline{n};\underline{n}')$
for a change from configuration $\underline{n}' \rightarrow \underline{n}$
are equal to the sum\\
$\begin{array}{rcl}
\displaystyle \sum_{k=1}^L \left\{ \delta_{n_k',0}\delta_{n_{k+1}',0} \left[
a_{21}\delta_{n_k,0}\delta_{n_{k+1},1} + a_{31}\delta_{n_k,1}\delta_{n_{k+1},0} + a_{41} \delta_{n_k,1}\delta_{n_{k+1},1} \right] + \right. & & \\
\delta_{n_k',0}\delta_{n_{k+1}',1}\left[
a_{12}\delta_{n_k,0}\delta_{n_{k+1},0} + a_{32}\delta_{n_k,1}\delta_{n_{k+1},0} + a_{42} \delta_{n_k,1}\delta_{n_{k+1},1} \right] + & & \\
\delta_{n_k',1}\delta_{n_{k+1}',0}\left[
a_{13}\delta_{n_k,0}\delta_{n_{k+1},0} + a_{23}\delta_{n_k,0}\delta_{n_{k+1},1} + a_{43} \delta_{n_k,1}\delta_{n_{k+1},1} \right] + & & \\
\left. \delta_{n_k',1}\delta_{n_{k+1}',1}\left[
a_{14}\delta_{n_k,0}\delta_{n_{k+1},0} + a_{24}\delta_{n_k,0}\delta_{n_{k+1},1} 
+ a_{34} \delta_{n_k,1}\delta_{n_{k+1},0} \right] \right\} . & & \\
 & & 
\end{array}$

This somewhat lengthy expression becomes more compact in the quantum
Hamiltonian formalism (\ref{1}): To each configuration 
$\underline{n}$ a vector $|\,\underline{n}\,\rangle$ which, together with the
transposed vectors $\langle\,\underline{n}\,|$, form an orthonormal basis
of $({\bf C}^2)^{\otimes L}$. In spin language this corresponds to a mapping to 
a spin 1/2 chain by identifying a vacancy (particle) at site $k$ with spin up 
(down) at this site. The probability distribution is then given by the
state vector $|\,f(t)\,\rangle=\sum_{\underline{n}} f(\underline{n};t) 
|\,\underline{n}\,\rangle$ and the formal solution of the master equation
(\ref{1}) in terms of the initial distribution $|\,f(0)\,\rangle$ is given
by $|\,f(t)\,\rangle = \exp(-Ht) |\,f(0)\,\rangle$. 
The stochastic dynamics are defined by the master equation (\ref{1}) 
with\cite{sch2}
\be
H = \sum_{k=1}^L h_k
\label{2}
\ee
where the matrices $h_k$ act non-trivially only on sites $k,k+1$ and are
given by
\be h_k = - \left( 
\begin{array}{cccc}
a_{11} & a_{12} & a_{13} & a_{14} \\
a_{21} & a_{22} & a_{23} & a_{24} \\
a_{31} & a_{32} & a_{33} & a_{34} \\
a_{41} & a_{42} & a_{43} & a_{44} 
\end{array} \right)_{k,k+1}
\label{3}
\ee
with $a_{jj}=-\sum_{\stackrel{i=1}{i\neq j}}^{4} a_{ij}$.

The connection of $H$ to the Heisenberg quantum chain becomes apparent by the 
similarity transformation $\tilde{H} = \Phi VHV^{-1}\Phi^{-1}$
with $V=\exp(S^+)$ where $S^+=\sum_{k=1}^L s^+_k$ and $s^{\pm}_k=(\sigma^x_k
\pm i\sigma^y_k)/2$ are the spin lowering and raising operators acting on site
$k$ and with $\Phi=\exp({\cal E}\sum_k k \sigma^z_k)$ where ${\cal E}$ is a 
suitably chosen constant\cite{sch2}. On the ten parameter submanifold defined by
\bea
a_{34} & = & a_{21} + a_{41} + a_{12} + a_{32} 
             - a_{23} - a_{43} - a_{14} \label{4} \\
a_{24} & = & a_{31} + a_{41} + a_{13} + a_{23} 
             - a_{32} - a_{42} - a_{14}.
\label{5}
\eea
the transition matrices have now the structure $\tilde{h_k} = 
h^{XXZ}_k + h^-_k$. Here $h^{XXZ}_k$ commutes with $S^z = \sum_{k=1}^L 
\sigma^z_k/2$ and $h^-_k$ is a sum of two parts which lower the $z$-component 
of the spin on sites $k,k+1$ by one and two units respectively. So one finds
\be
\tilde{H} = H^{XXZ} + H^-
\label{6}
\ee
where $H^{XXZ}$ is the Hamiltonian of the
anisotropic Heisenberg ferromagnet with twisted boundary conditions in a
magnetic field. The crucial observation is that $H^-$ {\em does not change 
the spectrum of} $H^{XXZ}$, since $H^{XXZ}$ may be block-diagonalized
into blocks with fixed quantum number $S^z$ and $H^-$ connects only blocks
of given $S^z$ with blocks with quantum numbers $S^z-1$ and 
$S^z-2$.\footnote{This mechanism was first noticed in a similar context in 
Alcaraz et al.\protect\cite{adhr}.} 

Quantities of interest are expectation values (i.e. $r$-point correlation functions) 
$\langle\,n_{k_1}(t) \dots n_{k_r}(t)\,\rangle_{f_0} =
\langle\,s\,|n_{k_1} \dots n_{k_r} e^{-Ht} |\,f(0)\, \rangle$ which give the
probability of finding particles on the set of sites $\{k_1, \dots , k_r\}$
at time $t$, if the initial distribution at time $t=0$ was $f_0$. 
Here $\langle \; s\,| = \sum_{\underline{n}} \langle\,\underline{n}\,|$
and $n_k = (1-\sigma^z_k)/2$ is the projector on states with a particle on site $k$.
From the Bethe ansatz one finds now that the spectrum has an energy gap (i.e. inverse 
correlation time) $\mu' = 4 a_{41} + 2(a_{21}+a_{31}) +a_{12}+a_{13} - a_{42} - a_{43} 
\geq 0$. If $\mu' = 0$ the dynamical exponent turns out to be $z=2$. 
Note also that $V$ transforms a $r$-point density correlation function
into a matrix element in the sector with $r$ down spins. Since
$H^-$ only creates down spins, only transformed initial states with
$l \leq r$ down spins will contribute to the correlation function.
This surprising
simplification allows for an exact calculation of the local average density
for any initial state even though we are dealing with a non-trivial 
interacting many particle system\cite{sch2}.

\section{The Dynamic Matrix Ansatz}

The results of the last section involve the constraints (\ref{4}),
(\ref{5}) and do not apply e.g. for the asymmetric exclusion process.
Also this model is integrable, but a calculation of time-dependent correlation 
functions has not yet been achieved. In order to solve this problem we now
formulate a dynamic matrix ansatz for the general reaction-diffusion system 
defined by (\ref{2}) and (\ref{3}), generalizing earlier work\cite{Sti1,Sti2} 
for diffusion only. Instead of periodic
boundary conditions we consider a system with open boundaries where
particles are injected (absorbed) at site 1 with rate $\alpha$ ($\gamma$)
and at site $L$ with rate $\delta$ ($\beta$). Therefore $H=b_1+b_L + 
\sum_{k=1}^{L-1}h_k$ with suitably chosen 
boundary matrices $b_1,b_L$\cite{Sti1}.

The ansatz is to take $|\,f(t)\,\rangle = \langle\langle\,W\,|\{ \prod_{k=1}^L 
(E(t) + D(t) \sigma^-_k) \} |\,0\,\rangle|\,V\,\rangle\rangle/Z_L$
where $|\,0\,\rangle$ is the state with all spins up and $D,E$ are 
time-dependent matrices satisfying an algebra obtained from the master equation 
(\ref{1}). The (time-independent) vectors $\langle\langle\,W\,|$ and $|\,V\;
\rangle\rangle$ on which $D$ and $E$ act are determined from the boundary
terms in the master equation and 
$Z_L=\langle\langle\,W\,|C^L|\,V\,\rangle\rangle$ where $C=D+E$ is a 
normalization.
In this framework the $r$-point density correlation function is given by
$\langle\,n_{k_1}(t) \dots n_{k_r}(t)\,\rangle_{f_0} =
\langle\langle\,W\,|C^{k_1-1}DC^{k_2-k_1-1}D\dots C^{L-k_r}|\,V\,\rangle\rangle/Z_L$.
Therefore, given a matrix representation of the algebra satisfied by
$D,E$, the computation of time-dependent correlation functions is reduced
to the much simpler calculation of matrix elements of a product of $L$
matrices.

It is easy to see that (\ref{1}) is solved if for each pair of sites one satisfies
\bea
& & (\frac{1}{2} \frac{d}{dt} + h_k)(E+D\sigma^-_k)(E+D\sigma^-_{k+1}) 
|\,0\,\rangle \;\; =
\nonumber \\
& & \left[(S+T\sigma^-_k)(E+D\sigma^-_{k+1}) - 
(E+D\sigma^-_k)(S+T\sigma^-_{k+1})\right] |\,0\,\rangle
\label{9}
\eea
where $S,T$ are auxiliary operators satisfying
\bea
\langle\langle\,W\,| \left[(\frac{1}{2} \frac{d}{dt} + b_1)(E+D\sigma^-_1) 
+(S+T\sigma^-_1) |\,0\,\rangle \right] & = & 0 \label{10a} \\
\left[ (\frac{1}{2} \frac{d}{dt} + b_L)(E+D\sigma^-_L) -(S+T\sigma^-_L) 
|\,0\,\rangle \right] |\,V\,\rangle\rangle & = & 0.
\label{10b}
\eea
By comparing each of the four terms in (\ref{9}) proportional to
$|\,0\,\rangle$, $\sigma^-_k|\,0\,\rangle$, $\sigma^-_{k+1}|\,0\,\rangle$
and $\sigma^-_k\sigma^-_{k+1}|\,0\,\rangle$ resp. one obtains four 
quadratic relations for the operators 
$D,E,S,T$. Eqs. (\ref{10a}) and (\ref{10b})
give two pairs of equations which define $\langle\langle\,W\,|$ 
and $|\,V\,\rangle\rangle$. Introducing
\bea
A^{(1)}_j & = & (a_{21}+a_{31}+a_{41}) E^2 - a_{12} ED
                 - a_{13} DE - a_{14} D^2  \\
B^{(1)}_j & = & - a_{21} E^2 + (a_{12}+a_{32}+a_{42}) ED
                 - a_{23} DE - a_{24} D^2 \\
B^{(2)}_j & = & - a_{31} E^2 - a_{32} ED
             + (a_{13}+a_{23}+a_{43}) DE - a_{34} D^2 \\
A^{(2)}_j & = & -a_{41} E^2 - a_{42} ED
             - a_{43} DE + (a_{14}+a_{24}+a_{34}) D^2.
\eea
one finds
\bea
\frac{1}{2}\frac{d}{dt} E^2 - [S,E]  & = & A^{(1)} \\
\frac{1}{2}\frac{d}{dt} ED - SD + ET & = & B^{(1)} \\
\frac{1}{2}\frac{d}{dt} DE - TE + DS & = & B^{(2)} \\
\frac{1}{2}\frac{d}{dt} D^2 - [T,D]  & = & A^{(2)} 
\eea
and
\bea
\langle\langle\,W\,|
\left\{\frac{1}{2}\frac{d}{dt} E - \alpha E + \gamma D + S \right\} & = & 0 \\
\langle\langle\,W\,|
\left\{\frac{1}{2}\frac{d}{dt} D + \alpha E - \gamma D + T \right\} & = & 0 \\
\left\{\frac{1}{2}\frac{d}{dt} E - \delta E + \beta D - S \right\} 
|\,V\,\rangle\rangle& = & 0 \\
\left\{\frac{1}{2}\frac{d}{dt} D + \delta E - \beta D - T \right\} 
|\,V\,\rangle\rangle& = & 0 .
\eea

One may reduce this algebra by assuming that $C$ is time-independent and has 
a representation where it is invertible. Eqs. (\ref{9}) then imply $[C,S+T] = 0$ 
and (\ref{10a}),(\ref{10b}) imply $\langle\langle\,W\,|(S+T)=0=(S+T)|\,V\,\rangle\rangle$. 
This can be solved by assuming $S+T=0$, which, as I would like to stress, is 
{\em not} the most general choice. Now one can express $S$ in terms of $C$ and 
$D$ and is left with only two further relations to be satisfied by $D$ and $C$
and two relations defining $\langle\langle\,W\,|$ and $|\,V\,\rangle\rangle$.
In particular, if (\ref{5}) and (\ref{6}) are satisfied, there is one relation 
involving $\dot{D}$ which is linear in $D$ and one relation quadratic in $D$.
For the symmetric exclusion model $a_{23}=a_{32}=1/2$ this dynamic algebra 
yields eigenvalue equations for the corresponding $XXX$-Hamiltonian with integrable,
but non-diagonal, symmetry breaking boundary fields\cite{Sti1,Sti2}. However, no 
matrix representation has been found yet. This raises the question whether 
non-trivial representations do exist at all. 

As I will show here for the first time, the answer to this question is yes, at 
least with some restrictions on the injection and absorption rates. Choosing a 
basis where $C$ is diagonal one finds the representation
\bea
C \; = \; \left(
\begin{array}{cc}
 1 & 0 \\
 0 & c
\end{array} \right) 
& , & 
D \; = \; \left(
\begin{array}{cc}
 d & \lambda e^{-\epsilon t} \\
 0 &  cd
\end{array} \right).
\label{15}
\eea 
with $\epsilon=(\alpha+\beta+\gamma+\delta)/2$, 
$c=1-\alpha-\gamma=(1-\beta-\delta)^{-1}$, $d=\alpha/(\alpha+\gamma)=
\delta/(\beta+\delta)$ and $\langle\langle\,W\,|$, $|\,V\,\rangle\rangle$ arbitrary 
but $\langle\langle\,W\,|\,V\,\rangle\rangle\neq 0$.
In this representation $\lambda$ is an arbitrary parameter specifying
the initial distribution. One may also use it for the construction of (right) 
eigenstates of $H$, since the expression $\langle\langle\,W\,|E^{k_1-1}
DE^{k_2-k_1-1}\dots E^{L-k_r}|\,V\,\rangle\rangle$ is a superposition
of wave functions $\Psi_{\epsilon_i}(k_1,\dots,k_r)$ of eigenstates with eigenvalues 
$\epsilon_i$. The argument is the position of $r$ down spins on 
sites $k_1,\dots,k_r$. Taking $\lambda=0$ corresponds to taking the stationary
distribution as initial state. This is an eigenstate with energy 0. The
terms proportional to $\lambda$ give the wave function for an eigenstate
with energy $\epsilon$. The quantity $1/(\ln{|c|})$ plays the role of a 
spatial correlation length.

\section{Conclusions}

I have shown that a 10-parameter class of stochastic 
reaction-diffusion systems can be mapped to a generalized Heisenberg
quantum chain, the spectrum of which can be obtained by the Bethe ansatz.
It turns out that time-dependent $r$-point density correlation functions
are given by the $l\leq r$-magnon sectors which allows for an explicit
calculation of correlators.
As an alternative to that approach a dynamic matrix ansatz was introduced for
the general 12-parameter model.
This ansatz reduces the calculation of all correlators to the calculation
of certain matrix elements. These matrices satisfy an algebra which is 
determined by the bulk dynamics of the process. The boundary conditions
determine which matrix elements one has to take.
In the case of the symmetric exclusion process
the algebra satisfied by the matrices can be used to obtain the spectrum
of the corresponding quantum Hamiltonian which is the isotropic
Heisenberg ferromagnet with non-diagonal boundary fields. An explicit 
time-dependent matrix representation was presented here for the first time. 
From this one explicitly obtains all $r$-point density correlators for a
one-parameter class of initial states. The corresponding eigenvectors of the
Heisenberg chain are the ground state with energy 0 and a bound state with
energy $\epsilon=(\alpha+\beta+\gamma+\delta)/2$.

The mapping in Sec.~2 relates a stochastic Hamiltonian to an integrable, 
non-stochastic Hamiltonian. It would be 
interesting to pursue this approach further and apply it to other systems.
Two of the most important unanswered questions are (i) the relationship
between the integrability of quantum chains and the dynamic matrix ansatz
and (ii) the existence and construction of representations of the 
general reaction-diffusion algebra derived in Sec.~3.

\section*{Acknowledgments}
It is a pleasure to acknowledge fruitful discussions with V. Hakim, V.
Rittenberg and R.B. Stinchcombe and to thank the organizers of this conference
for their kind invitation. This work was supported by an EEC fellowship
under the Human Capital and Mobility program.

\end{document}